\documentclass[conference]{IEEEtran}
\IEEEoverridecommandlockouts
\usepackage{cite}

\usepackage{amsmath,amssymb,amsfonts}
\usepackage{algorithmic}
\usepackage{graphicx}
\newcommand*{\rom}[1]{\expandafter\@slowromancap\romannumeral #1@}
\usepackage{textcomp}
\usepackage{tabularx}
\usepackage{booktabs}
\usepackage{bm}
\usepackage{babel}

\usepackage{amsmath}
\usepackage{xcolor}
\def\BibTeX{{\rm B\kern-.05em{\sc i\kern-.025em b}\kern-.08em
    T\kern-.1667em\lower.7ex\hbox{E}\kern-.125emX}}

\usepackage[autostyle=true]{csquotes}

\begin{document}

\title{A Deep Learning Based Decoder for Concatenated Coding over Deletion Channels\\
\thanks{This work was funded by the European Union through the ERC Advanced Grant 101054904: TRANCIDS. Views and opinions expressed are, however, those of the authors only and do not necessarily reflect those of the European Union or the European Research Council Executive Agency. Neither the European Union nor the granting authority can be held responsible for them.}}

\author{\IEEEauthorblockN{E. Uras Kargı\textsuperscript{$\dagger$}\thanks{\textsuperscript{$\dagger$}E. Uras Kargı's work is also supported by Vodafone within the framework of the 5G and Beyond Joint Graduate Support Program coordinated by the Information and Communication Technologies Authority.},
Tolga M. Duman}
\IEEEauthorblockA{Dept. of Electrical and Electronics Engineering, Bilkent University \\
Email: \{kargi, duman\}@ee.bilkent.edu.tr}}
\maketitle

\begin{abstract}

In this paper, we introduce a deep learning-based decoder designed for concatenated coding schemes over a deletion/substitution channel. Specifically, we focus on serially concatenated codes, where the outer code is either a convolutional or a low-density parity-check (LDPC) code, and the inner code is a marker code. We utilize Bidirectional Gated Recurrent Units (BI-GRUs) as log-likelihood ratio (LLR) estimators and outer code decoders for estimating the message bits. Our results indicate that decoders powered by BI-GRUs perform comparably in terms of error rates with the MAP detection of the marker code. We also find that a single network can work well for a wide range of channel parameters. In addition, it is possible to use a single BI-GRU based network to estimate the message bits via one-shot decoding when the outer code is a convolutional code. \footnote{Code is available at https://github.com/Bilkent-CTAR-Lab/DNN-for-Deletion-Channel}
\end{abstract}


\section{Introduction}

We consider binary-input binary-output channels with synchronization errors modeled as deletions and substitutions. Such channels are encountered in various real-world scenarios, such as storage systems, recording channels, and wireless communications \cite{recording}. Another scenario in which these models are observed is DNA storage, where existing nucleotides are deleted or new ones are inserted into the DNA strands \cite{dna}.

Dobrushin proved that channels with independent and identically distributed (i.i.d.) synchronization errors are information stable; hence, their Shannon capacity exists \cite{capacity}. While channel capacity calculation remained elusive despite many efforts over the last six decades, computable upper and lower bounds have been developed, e.g., \cite{bound1, bound2, bound3}. There are also practical channel codes designed for such channels, including algebraic codes, convolutional codes, and serially concatenated codes, e.g., \cite{alg2, alg1, conv2}.

An important class of codes for insertion/deletion channels is obtained by concatenating an outer code with an inner marker or watermark code \cite{marker1, water, marker3}. Markers are specific sequences of bits inserted into the data stream at regular intervals. They serve as reference points for maintaining synchronization, allowing the receiver to detect if there are insertions or deletions between the markers. The outer code can be a powerful channel code such as an LDPC code, a polar code, or a convolutional code \cite{marker1, polar1}. On the other hand, watermark codes employ patterns within a sparsely encoded sequence of bits to recognize and recover from insertion and deletion errors and can be efficiently used with non-binary LDPC codes \cite{water}.

MAP detection is used to identify the synchronization errors by estimating the log-likelihood ratios of the transmitted bits, which are then input to the outer decoder for serial concatenation of an outer code with an inner code over channels with insertions and deletions \cite{water,marker1}. As an alternative, we argue that recurrent deep learning architectures may also be promising candidates to deliver a similar performance to that of the MAP detector while offering further benefits. Hence, they represent a potential avenue for development of new decoders for channels with insertions and deletions within the context of marker codes.

Various deep-learning architectures have been developed for coding over different channels. Several decoder architectures based on multi-layer perceptrons (MLP) have been proposed for one-shot decoding of channel codes in \cite{linear1, on1}. Decoders based on MLPs have demonstrated that structured codes are learnable by neural networks, and generalizations to unseen codewords are possible. However, for large code lengths, decoding via MLPs does not appear feasible \cite{on1}. Recurrent Neural Networks (RNN) have shown excellent performance in handling time-series data due to their internal memory. RNN variants, namely gated recurrent units (GRU) \cite{gru} and long short-term memory (LSTM) \cite{lstm}, have further enhanced the performance of RNNs by alleviating the vanishing gradient problem and allowing the network to investigate further the relationship between the data separated by large time steps. As a result, networks based on LSTM and GRU architectures are proposed for decoding convolutional codes \cite{rnn based decoder}, demonstrating a close to optimal error performance obtained by MAP decoders. GRUs have also been employed to decode turbo codes, replacing the BCJR blocks in iterative decoding \cite{rnn based decoder 2}. Moreover, decoders based on the transformer architecture have been demonstrated successfully for different channel codes by combining the attention mechanism and the code structure \cite{code transformer}.

In this paper, we develop a deep learning (DL) based decoder for concatenation of an outer LDPC or convolutional code with an inner marker code used over channels with deletions and substitution errors. Our contributions are to develop:

\begin{itemize}
    \item a BI-GRU estimator for the bit LLRs for the inner marker code,
    \item one-shot decoder for serial concatenation of convolutional and marker codes via a complete BI-GRU based architecture.
\end{itemize}

Numerical results demonstrate that the newly developed DNN-based decoders offer robust decoding over a deletion channel by replacing the main parts of the decoder.

We note that after the original submission of this paper, a closely related paper on deep learning-based decoders for coding over insertion/deletion channels was uploaded on Arxiv \cite{dl_ins_del}. The authors propose two distinct setups: a model-based configuration within the framework of forward-backward equations and a model-free setup, where the implementation focuses on end-to-end estimation of LLRs within a framework akin to ours. The differences in our work and \cite{dl_ins_del} can be highlighted as the incorporation of one-shot decoding approach applied to the receiver in our work, distinct preprocessing steps applied to the received bits, and variations in training methodologies (for instance, utilization of a static dataset in [20] versus simulation-based generation of mini-batches during training in our approach). 

The paper is organized as follows. Section \ref{system} discusses the system model. Section \ref{setup} describes the decoding scheme employing the proposed BI-GRU architecture. Section \ref{results} presents some numerical examples demonstrating the performance of the proposed architectures. Finally, the paper is concluded in Section \ref{conc}.

\section{System Model}
\label{system}
We consider transmission over binary channels susceptible to deletions and bit flips, modeling synchronization errors. Let $P_d$ and $P_s$ be the deletion and substitution probability of the channel, respectively. That is, each bit is either deleted with probability $P_d$ or transmitted incorrectly with probability $(1-P_d)P_s$ or transmitted correctly with probability $(1-P_d)(1-P_s)$. The deletions and substitutions are independent for different uses of the channel.

\begin{figure}[t] 
\centering
\includegraphics[scale = 0.55]{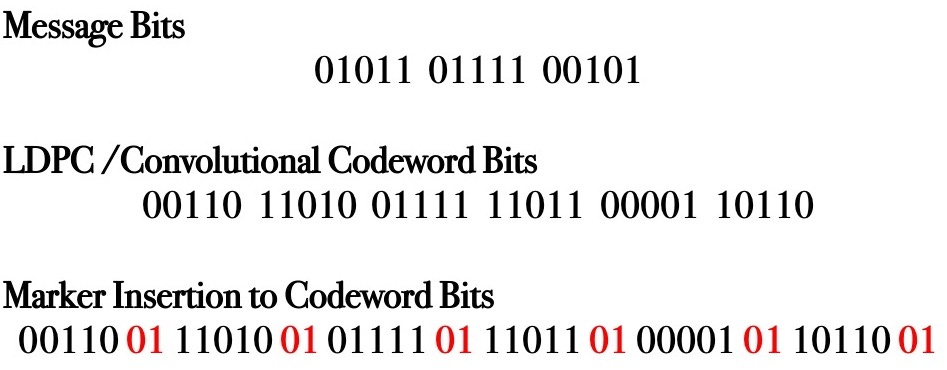}
\caption{Insertion of marker bits into the LDPC or convolutional coded bits. This case is for when $N_c = 5$, $N_m = 2$, $k/n = 0.5$, $r = 5/14$ and the two-bit marker of $[0,1]$.}
\label{ilus}
\end{figure}

We adopt a serially concatenated scheme as the specific channel coding approach in which the outer code is either an LDPC or a convolutional code, and the inner code is a marker code. Message vector \(\bm{m} = (m_1,\ldots,m_k)\) of length $k$ where $m_t \in \{0,1\}$ is encoded via the outer code, and then the encoded sequence $\bm{c} = (c_1,\ldots,c_n)$ of length $n$ where $c_t \in \{0,1\}$ is interleaved, resulting in $\bm{c}^{\pi}$. The rate of outer code is $k/n$. Then, a pre-selected marker sequence with $N_m$ bits is inserted after every $N_c$ coded bits in $\bm{c}^{\pi}$. The rate of the overall code is $r = \frac{N_c k}{(N_c + N_m) n}$. The resulting sequence of $\frac{n(N_c + N_m)}{N_c}$ bits comprise the overall codeword as illustrated in Fig. \ref{ilus}. 

The codeword is transmitted over a deletion/substitution channel. At the receiver, a MAP detector can be used to estimate the LLRs of the bits encoded by the outer code, and the obtained LLRs can be input into the outer code's decoder. In this paper, our objective is to develop DNN-based alternatives for the MAP detector and the overall decoder for the case of convolutional codes concatenated with marker codes.

\section{Proposed Decoding Scheme Employing BI-GRUs}
\label{setup}
\begin{figure}[t]

\centering
\includegraphics[scale = 0.45]{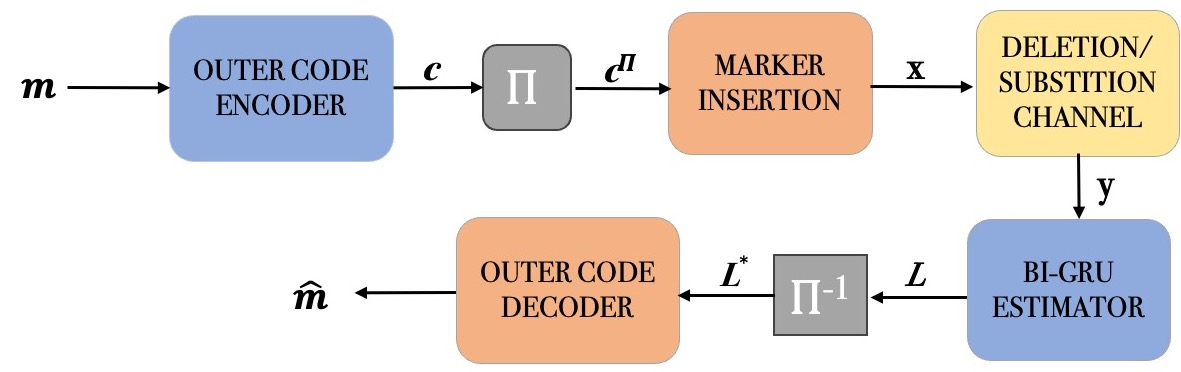}
\caption{The block diagram for the first setup.}
\label{setup1}
\end{figure}

\begin{figure}[t]
\centering
\includegraphics[scale = 0.44]{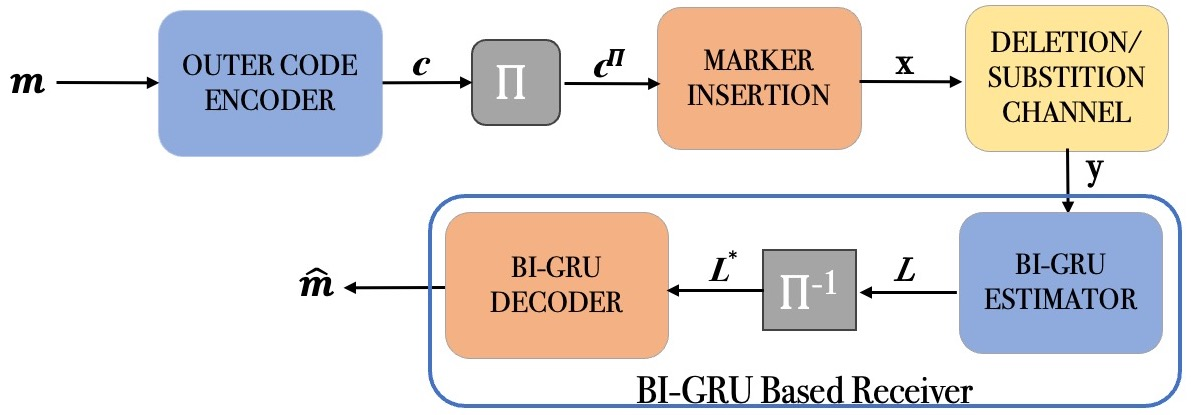}
\caption{The block diagram for the second setup.}
\label{setup2}
\end{figure}

We consider two different setups. In the first setup, we have a serial concatenation of an LDPC (or a convolutional code) with a marker code, and a BI-GRU architecture is developed to estimate the LLRs of the bits input to the marker code. Fig. \ref{setup1} illustrates the block diagram of the first setup. In the second setup, we focus on the serial concatenation of a convolutional code with a marker code and develop BI-GRU architectures that perform one-shot decoding. That is, two BI-GRU architectures are combined to estimate the LLRs and decode the message bits in one step. Fig. \ref{setup2} depicts the block diagram of the second setup, where the entire receiver is replaced by BI-GRU architectures.

The transmitted sequence of length $T$ is denoted by $\bm{x} = (x_1,\ldots, x_T)$ and the received sequence by $\bm{y} = (y_1,\ldots, y_R)$ where the length of the received sequence, $R$, is a random variable due to the deletions, $R \leq T$. The proposed BI-GRU based architecture estimates the LLR of each transmitted bit, denoted by $\bm{L} = (L_1, \ldots, L_T)$. The estimated LLRs are first de-interleaved, and then those corresponding to marker bits are removed, resulting in the vector $\bm{L}^{*} = (L_1,\ldots, L_n) $. $\bm{L}^{*}$ is passed to the outer code decoder. A sum-product algorithm-based decoder can be used for LDPC codes, or the Viterbi algorithm can be employed for convolutional codes to estimate the message vector \(\hat{\bm{m}} = (\hat{m}_1,\ldots \hat{m}_k)\). Other alternatives for the outer code decoder are also possible. For instance, we employ a BI-GRU architecture to estimate the message vector for the second setup \cite{rnn based decoder}. 

\subsection{Bi-directional RNN}
\label{birnn_txt}

 We choose the BI-RNN architecture since it can handle variable-length sequences produced at the output of a deletion channel, and it resembles the structure of the MAP detector realized by the forward-backward equations. BI-RNN consists of two independent RNN architectures, forward and backward RNN, enabling the network to capture information effectively in both forward and backward directions. In a forward RNN, the hidden state at a given time \( t\), \(\overrightarrow{h_t}\) represents the prior information. Let \(\overrightarrow{W}_{rec}\) and \( \overrightarrow{W}_{in}\) denote the recurrent weight matrix and the input weight matrix of forward RNN, respectively, and $f$ denote an activation function. We can compute \(\overrightarrow{h_t}\) as follows:
\begin{equation}
\label{forward_RNN}
\overrightarrow{h_t} = f(\overrightarrow{W}_{rec} \overrightarrow{h}_{t-1} + \overrightarrow{W}_{in} y_t).
\end{equation}
In a backward RNN, the hidden state at a given time \( t\) represents the information about the sequence to the right of the current input, \( \overleftarrow{h_t}\). Let \(\overleftarrow{W}_{rec}\) and \(\overleftarrow{W}_{in}\) denote the recurrent weight matrix and the input weight matrix of backward RNN, respectively. We can compute \(\overleftarrow{h_t}\) as follows:
\begin{equation}
\label{backward_RNN}
\overleftarrow{h_t} = f(\overleftarrow{W}_{rec} \overleftarrow{h}_{t+1} + \overleftarrow{W}_{in} y_t).
\end{equation}
A layer of BI-RNN \cite{bi-rnn} combines a forward and a backward RNN described above. A linear layer with a sigmoid activation function is added on top of the final BI-RNN layer to calculate the conditional probability of $x_t = 1$ given $\bm{y}$. Let \(W_{o}\) denote the output weight matrix of the network, \( \sigma\) be the sigmoid function and $[\: ;\:]$ be the concatenation operator. We calculate the output of the network as $P(x_t = 1|\bm{y}) = \sigma(W_{o} [\overrightarrow{h_t} ; \overleftarrow{h_t}])$. Then, the LLR of $x_t$, $L(x_t| \bm{y})$ for \( t= 1,2,\ldots, T\) is computed using:
\begin{flalign}
\label{out}
\begin{split}
L(x_t| \bm{y})   \triangleq  \text{log} \frac{P(x_t = 0| \bm{y}) }{P(x_t = 1|\bm{y})} = \text{log} \frac{1 - \sigma(W_{o} [\overrightarrow{h_t} ; \overleftarrow{h_t}])}{\sigma(W_{o} [\overrightarrow{h_t} ; \overleftarrow{h_t}])}.
\end{split}
\end{flalign}
\subsection{BI-GRU as an LLR Estimator}

We implement the BI-GRU architecture as an LLR estimator for the marker code. GRU and LSTM networks address the issues of vanishing and exploding gradients commonly faced in RNNs, while also significantly enhancing memory capabilities compared to simple RNNs \cite{gru,lstm, vanish}. GRUs are chosen instead of LSTMs since the latter is more challenging to train. The BI-GRU is an extension of BI-RNN and has two independent components that are similar to BI-RNN outlined in Section \ref{birnn_txt}, namely forward and backward BI-GRU. Let $\odot$ denote the Hadamard Product; equations of a forward GRU layer are as follows \cite{gru}:
\begin{flalign}
\label{gru}
\begin{split}
& z_t = \sigma(W_z  [h_{t-1} ; y_t])) \\
& r_t = \sigma(W_r  [h_{t-1} ; y_t])) \\
& \tilde{h_t} = \text{tanh}(W_h [r_t \odot h_{t-1} ; y_t]) \\ 
& h_t = (1-z_t) \odot h_{t-1} + z_t\odot\tilde{h_t},
\end{split}
\end{flalign}
\noindent where $z_t$ is the vector that controls how much new information is added to the hidden state, $\tilde{h_t}$ is the vector that represents the new candidate information, and $r_t$ is the vector that controls how much the previous state information is passed to the temporary variable $\tilde{h_t}$. All of $z_t$, $r_t$, $\tilde{h_t}$, and $h_t$ have a size of $d_{\text{BI-GRU}}$.

Similarly, backward GRU equations are easily obtained as in ($\ref{backward_RNN}$). If a BI-GRU architecture with $N_{\text{BI-GRU}}$ layers is employed, then $\overrightarrow{h_t}$ and $\overleftarrow{h_t}$ serve as inputs to the forward and backward GRU of the following layer, respectively. An MLP with $N_{\text{MLP}}$ layers with dimensions $d_{\text{MLP}}$ is added at the top of the final layer of BI-GRU, and then (\ref{out}) can be used to estimate the bit LLR. 

\subsection{One Shot Decoding of Convolutional Codes Concatenated with Markers} \label{subsec_conv_only}

As an alternative approach, we utilize the BI-GRU architecture as the decoder for the outer code in the second setup, employed in \cite{rnn based decoder}. At time step $t$, two consecutive LLRs $(L^{*}_{2t-1}$, $L^{*}_{2t})$ of the vector $\bm{L^{*}}$ are input to the BI-GRU network and the output is the message estimate $\hat{m_t}$ for \(t  = 1,\ldots,k\). A linear layer with a sigmoid activation function is added on top of the final BI-GRU layer to estimate message bits. 


\subsection{Implementation Details} \label{sec_training}

We now describe the training procedure of the BI-GRU architecture for the proposed setups. The BI-GRU networks are trained with a supervised learning approach: For the BI-GRU based estimator, labels are the original transmitted sequences \( \bm{x}\) over a deletion/substitution channel, and the inputs are the received sequence $\bm{y}$ transformed by $-2\bm{y} + 1$ whereas for the BI-GRU decoder, labels are message sequences \( m\) and inputs are the LLR vectors $\bm{L^*}$ obtained from the BI-GRU based estimator. We do not use a pre-determined data set, instead, we generate each mini-batch randomly during training. I.e, we generate random message bits $\bm{m}$ (labels), encode them using an outer encoder to produce $\bm{x}$, which are then transmitted through a deletion/substitution channel. This process yields either $\bm{y}$ (inputs to the estimator) or $\bm{L^*}$ (inputs to the decoder), obtained by estimating LLRs from $\bm{y}$. Note that, the BI-GRU based decoders and estimators are trained independently. 

 We use the Adam optimizer \cite{adam} and a staircase learning decay \cite{lrdecay}, for which in each $S$ step of the Adam optimizer, the learning rate decays exponentially with parameter $D$ during training. If not stated, we use the default hyperparameters of the Adam optimizer. Batch normalization layers are added between the BI-GRU layers to train the network smoothly for both the estimator and the decoder \cite{batch}.

\section{Numerical Results}
\label{results}



 As a first example, we use a regular, rate 1/2 LDPC code with $n = 204$ and $k = 102$ (taken from \cite{mackay}). The maximum iteration number of the sum-product decoder is set to $100$. We use a two-bit marker \([0,1]\) and choose \(N_c\) as 5 and 10, which results in code rates of \(0.358\) and \(0.4167\), respectively. We train two networks for \(30,000\) steps, fixing \(P_d=0.05\) and $P_s=0$. The initial learning rate is chosen as $9\cdot10^{-5}$ for the Adam optimizer, and we use a mini-batch size of $16$. The overall input length is $T$, and since $R \leq T$, a total of $T-R$ zeros are padded for training and inference. From our numerical experiments, two consecutive bits after $t$ including the current one, \( (y_t, y_{t+1})\) are input to the BI-GRU at time step $t$ since it performs better than giving only $y_t$ as an input. We set \(D=0.95\) and, \(S=1000\) as detailed in Section \ref{sec_training}. We use gradient clipping with $0.1$ to avoid the vanishing gradient problem. LLRs obtained from the BI-GRU estimator are clipped in between $[-10,10]$ during testing. The loss function is chosen as mean-squared error (MSE). Parameters of the trained BI-GRUs are given in the first and second columns in Table \ref{table1}.

\begin{figure}
\centering
\includegraphics[trim={0cm 0 0 20},clip,width=0.5\textwidth]{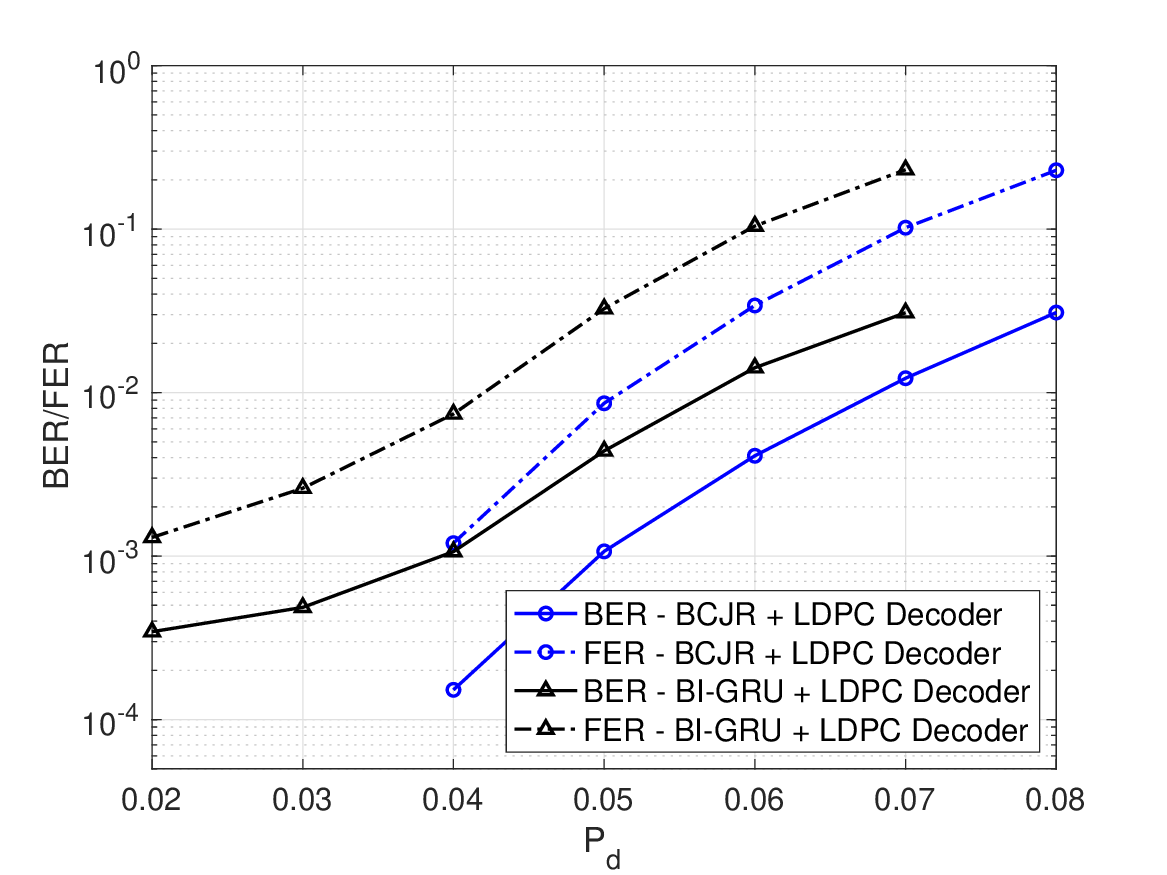}
\caption{BER/FER as a function of the deletion probability for the first setup with BI-GRU as an estimator when $N_c = 5$.}  
\label{result1}
\end{figure}

\begin{figure}
\centering
\includegraphics[trim={0cm 0 0 20},clip,width=0.5\textwidth]{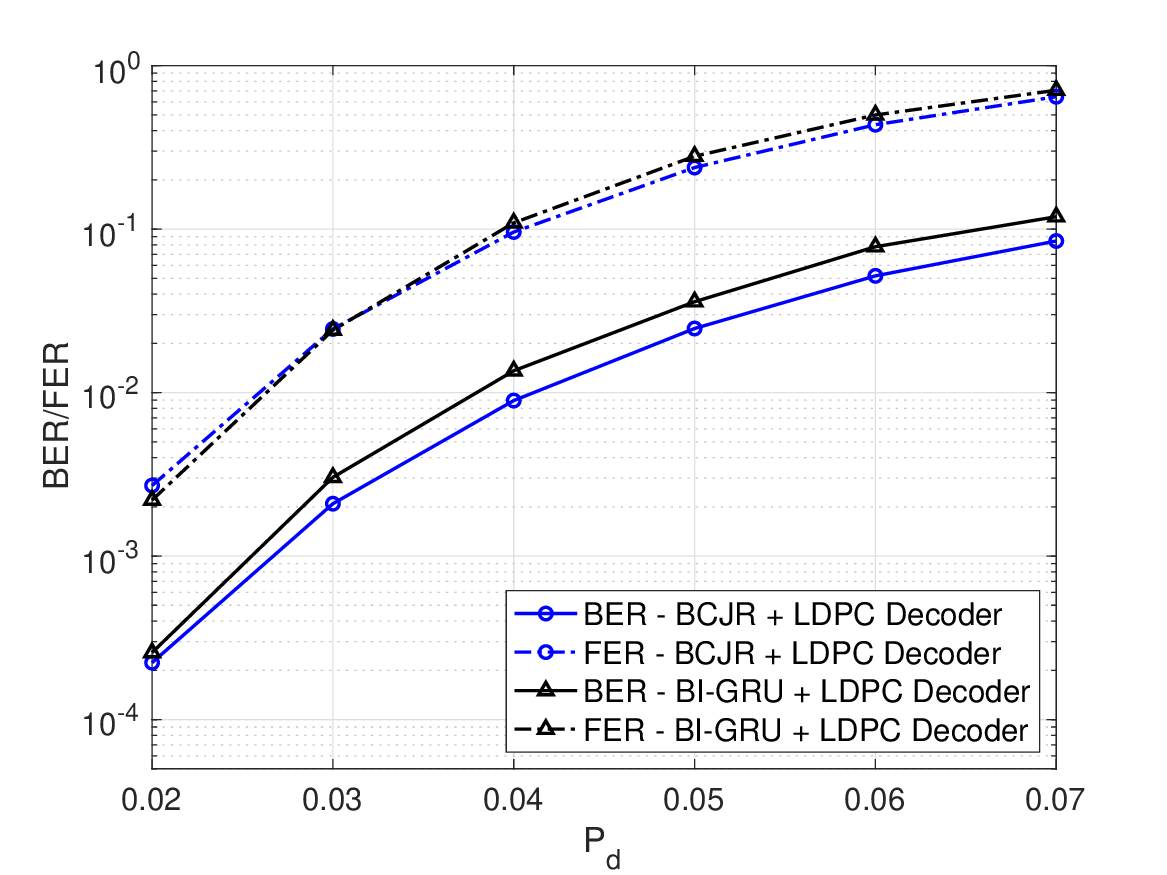}
\caption{BER/FER as a function of the deletion probability for the first setup with BI-GRU as an estimator when $N_c = 10$..}  
\label{result2}
\end{figure}

Fig. \ref{result1} and Fig. \ref{result2} show the error rate performance of the proposed decoders for the first setup. We observe that the proposed decoders generalize well over the deletion probabilities in the range of $0.02-0.07$ despite being trained only with $P_d= 0.05$. The performance gap between the baseline and proposed decoder is smaller for $N_c = 10$. For instance, when \(N_c = 5\), a BER of \( 10^{-3} \) is achieved at a deletion probability of $0.04$ for the proposed decoder and $0.05$ for the baseline decoder. When \(N_c = 10\), the same BER is obtained at a deletion probability of $0.025$ for the proposed decoder and at around $0.027$ for the baseline. 

As a second example, we trained a network to be robust to different deletion and substitution probabilities. We use the same LDPC code and the marker sequence as in the first example and choose \(N_c = 10\). The network is trained by mixing deletion probabilities in $[0.01, 0.1]$ with $0.01$ increments and mixing substitution probabilities in $[0, 0.1]$ with $0.01$ increments. In this case, unlike the first approach, the bits before timestep $t$, including the current one, \( (y_1, \ldots, y_t)\), are input to the BI-GRU at time step $t$. The initial learning rate is chosen as $9\cdot10^{-4}$, and we use a mini-batch size of $16$. The loss function is chosen as binary cross entropy (BCE). In this case, since the input size is increased, we use smaller models. Parameters of the robust BI-GRU are given in the third column in Table \ref{table1}.
\begin{table}[h] 
\centering
\begin{tabular}{cccccc}
\toprule
& \multicolumn{3}{c}{Estimators} & \multicolumn{2}{c}{Decoder} \\
\cmidrule(lr){2-4} \cmidrule(lr){5-6}
& Estimator 1 & Estimator 2 & Estimator 3 &  \\
\midrule
$N_c$ & 5 & 10 & 10 & - \\
$N_{\text{BI-GRU}}$ & 6 & 8 & 4 & 2 \\
$d_{\text{BI-GRU}}$ & 1024 & 1024 & 128 & 400 \\
$N_{\text{MLP}}$ & 4 & 4 & 3 & 2 \\
$d_{\text{MLP}}$ & [256,128,32,1] & [256,128,32,1] & [128, 32, 1] & [32, 1]\\
\bottomrule
\end{tabular}
\vspace{5pt}
\caption{Parameters of trained BI-GRUs as an estimator or a decoder.}
\label{table1}
\end{table}

\begin{figure}
\centering
\includegraphics[trim={0cm 0 0 20},clip,width=0.5\textwidth]{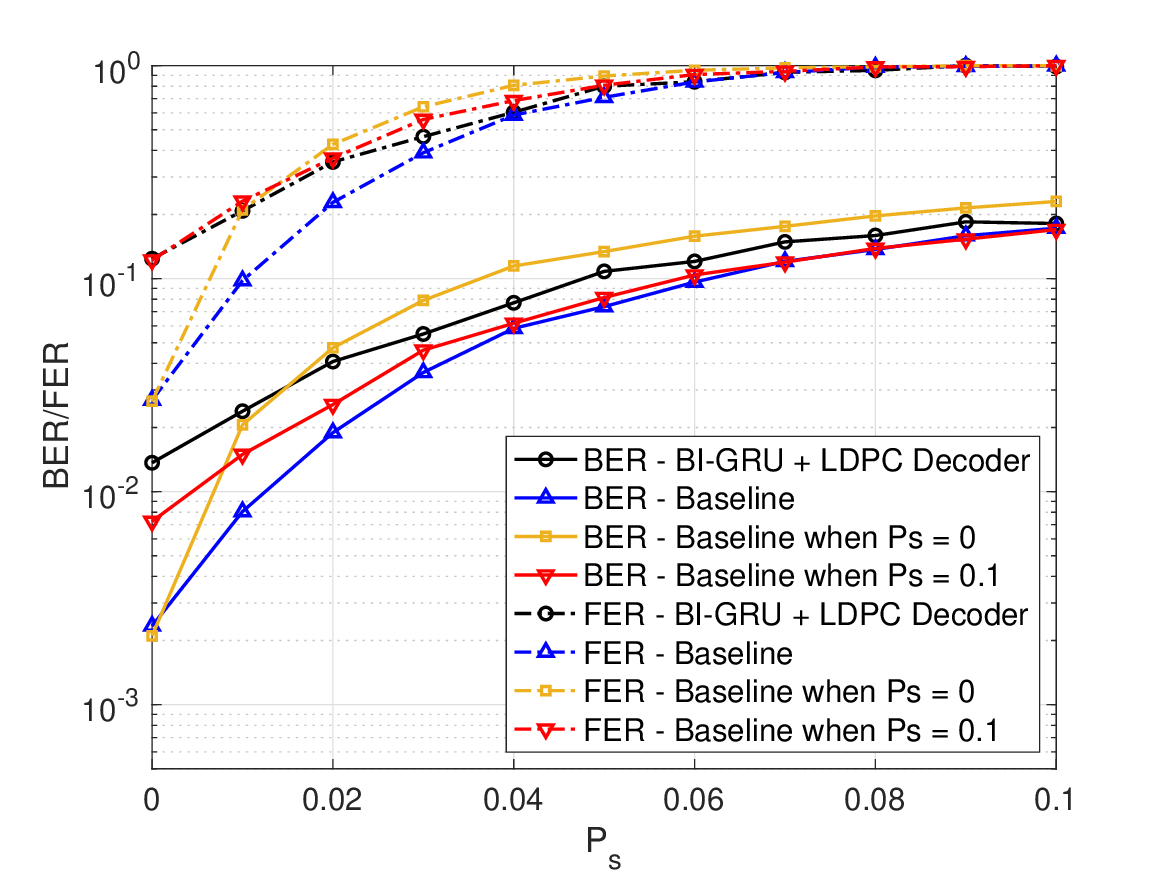}
\caption{BER/FER as a function of substitution probability for a channel with $P_d = 0.03$.}  \label{result_robust1}
\end{figure}

\begin{figure}
\centering
\includegraphics[trim={0cm 0 0 20},clip,width=0.5\textwidth]{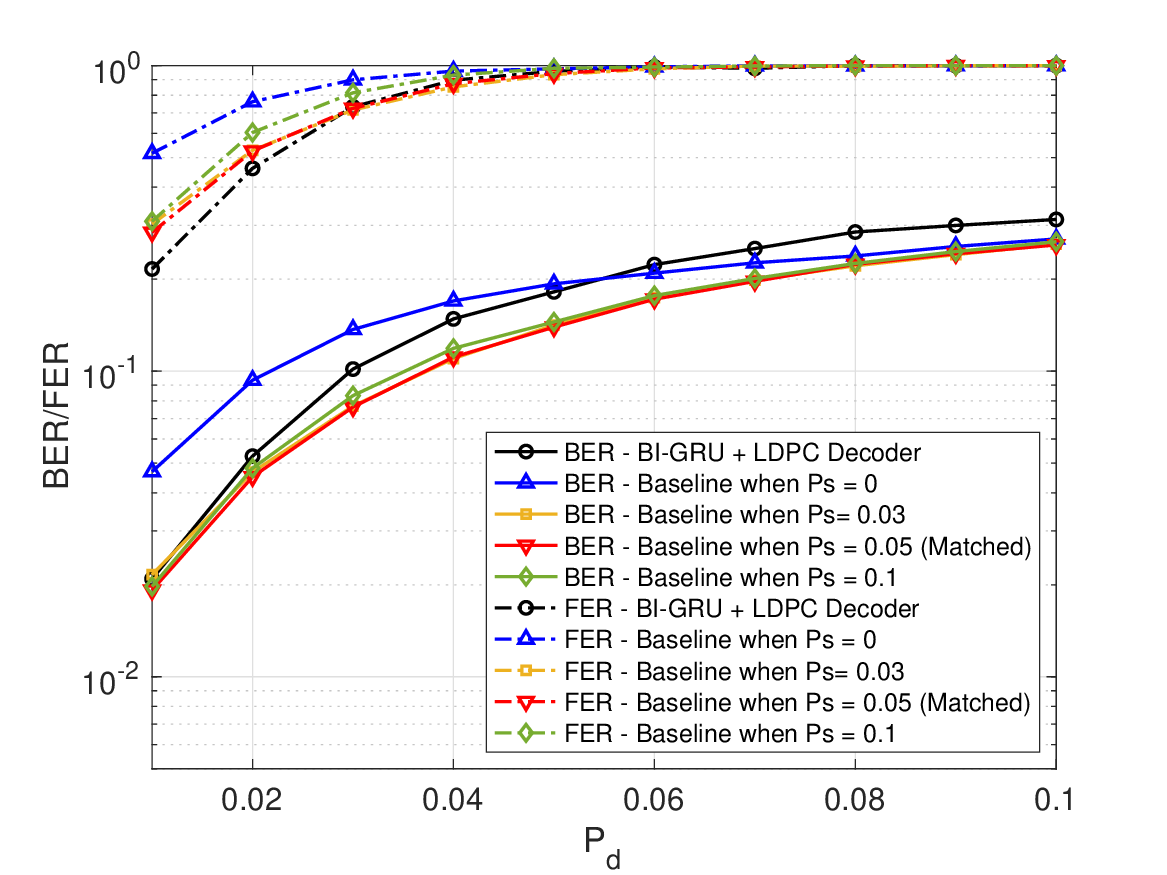}
\caption{BER/FER as a function of deletion probability for a channel with $P_s = 0.05$.}  
\label{result_robust2}
\end{figure}

An important point is that the baseline approach employing the BCJR algorithm uses the deletion and substitution probabilities of the channel to calculate the LLR estimates, while the BI-GRU based estimator does not. Fig. \ref{result_robust1} compares the performance of the proposed decoder with the baseline over a range of substitution probabilities for a deletion/substitution channel with $P_d = 0.03$, and Fig. \ref{result_robust2} compares the performance of the proposed decoder with the baseline over a range of deletion probabilities when $P_s = 0.05$. For both figures, we assume that the channel parameters are not known accurately at the receiver. The baseline decoder estimates the deletion probability using \( \hat{P_d} = \frac{T-R}{T}\). Four different results are produced by taking the substitution probabilities used at the receiver as \{0, 0.03, 0.05, 0.1\}. We observe that the proposed BI-GRU based decoder exhibits improved performance in some cases. Specifically, when the baseline decoder uses a substitution probability of \(0\), for a range of deletion probabilities, the proposed decoder outperforms the BCJR algorithm results. Notably, in terms of the FER performance, the proposed decoders outperform the mismatched decoder for a wider range of channel parameters.

We also provide an example where we replace the decoder of the outer convolutional code with the fully BI-GRU-based decoder, detailed in Section \ref{subsec_conv_only}. We use a rate $1/2$, $(5,7)_{octal}$ convolutional code with $n=210$, and $k=105$. The parameters of the BI-GRU decoder for the convolutional code are given in the fourth column of Table \ref{table1}. We train the BI-GRU decoder for $5000$ steps with an initial learning rate of $3\cdot10^{-4}$. We do not apply any learning rate decay for training in this case. We use the same BI-GRU estimator previously used for the first setup, whose parameters are given in the second column of Table \ref{table1}. We consider two outer decoders as baselines: Viterbi algorithm with soft-decision decoding (SDD) and with hard-decision decoding (HDD). For the SDD case, LLRs of the BCJR algorithm-based estimator are used as if they are the outputs of an additive white Gaussian noise (AWGN) channel, and a correlation metric is used. For the case of the Viterbi algorithm with HDD, the LLRs of the estimator (it can be BI-GRU based or BCJR-algorithm based) are mapped to hard decision values: $0$ if $L(x_t = 0| \bm{y}) \geq 0$ and $1$ if $L(x_t = 0| \bm{y}) < 0$, which are then input to the Viterbi algorithm.

\begin{figure}
\centering
\includegraphics[trim={0cm 0 0 20},clip,width=0.5\textwidth]{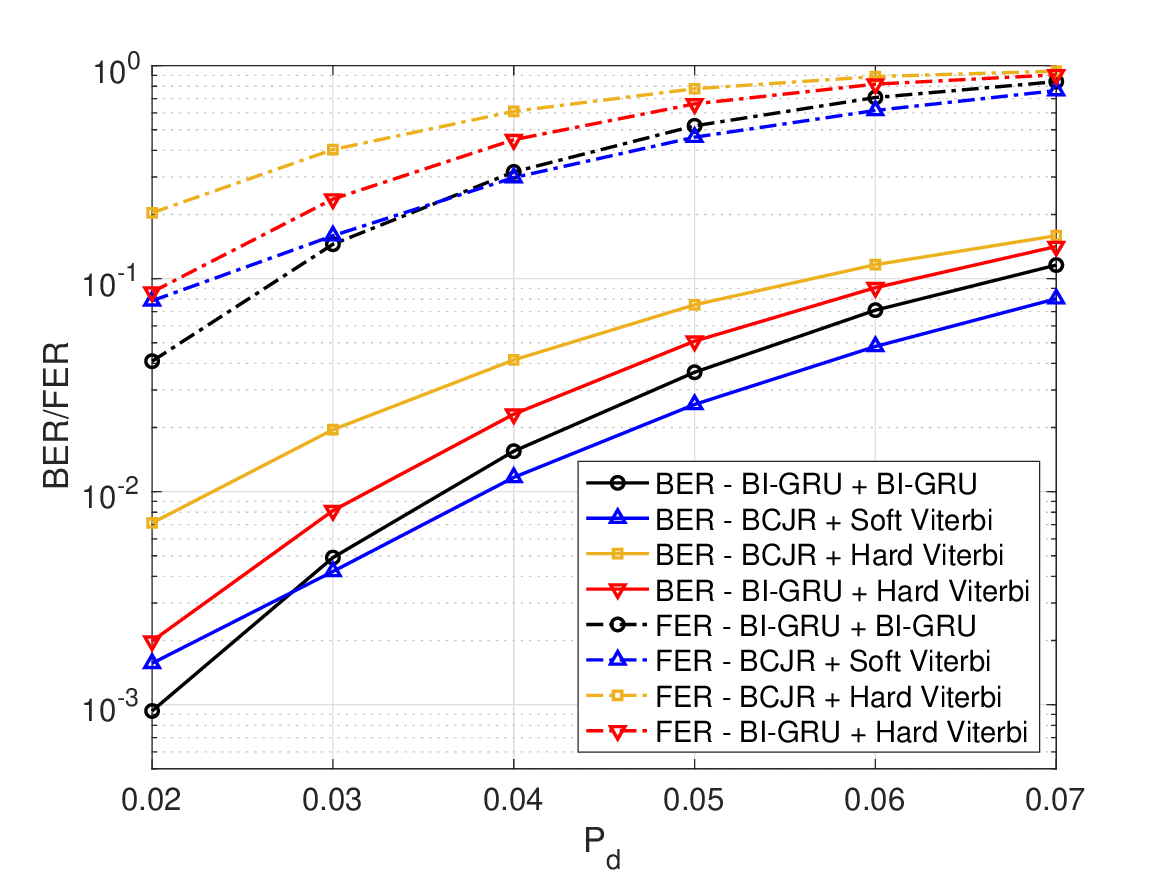}
\caption{BER/FER as a function of deletion probability (with no substitution errors).}  
\label{result_conv}
\end{figure}

Fig. \ref{result_conv} depicts the resulting error rates. We observe that the fully BI-GRU based decoder performs similarly to the baseline decoder, when the Viterbi algorithm with SDD is used as the outer decoder and a BCJR algorithm is used as the estimator. Given that the training objective of the BI-GRU estimator is binary classification, there may be enhanced compatibility with the Viterbi algorithm using HDD; the decoder using the BI-GRU estimator along with Viterbi algorithm employing HDD performs better than the baseline approach paired with the same outer decoding algorithm for all the deletion probabilities considered. 

Finally, we want to comment on the numerical results of the recent work \cite{dl_ins_del}. We find that the results in \cite{dl_ins_del} are compatible with ours: the BI-GRU based estimators perform similarly to our trained decoders in terms of their performance difference with the baseline decoders realized using forward-backward equations. The model-based decoders in \cite{dl_ins_del} utilizing a forward-backward algorithm aimed to estimate $P_d, P_s$, and $P_i$, perform better than the BCJR-algorithm based decoders when the channel parameters are not known to the recevier. The authors also experiment with the burst insertion/deletion channels of which the deletions and insertions are not independent for different uses and show that without knowing the channel model, one may develop BI-GRU based decoders and argue that their performance in terms of error rates is superior to those of the BCJR-based decoders.

\section{Conclusions}
\label{conc}
This paper develops a BI-GRU architecture as a decoder for concatenated codes used over deletion channels. This architecture is chosen since it can handle variable-length inputs making it suitable for processing the received sequences over channels with symbol losses. The BI-GRU based deep neural network is trained to estimate the LLRs of the bits input to the marker code, which are then decoded using the outer code constraints. Furthermore, a one-shot solution is developed for the case with the outer convolutional code. The results reveal that the newly developed deep-learning based decoders offer competitive performance with the previously employed solutions, and they are robust. In some cases, they outperform the results obtained with the BCJR algorithm when there is a mismatch between the actual channel parameters and those used at the receiver. The current implementation of the BI-GRU network is demanding; however, different approaches, such as pruning and quantization, can be employed to reduce its complexity. 

Many promising research directions can be identified. For instance, deep-learning-based decoders can be developed for more general channel models with synchronization errors, e.g., those including insertions along with deletions and even intersymbol interference and AWGN (motivated by the channel models in bit-patterned media recording technologies \cite{media2} or nanopore sequencing in DNA storage \cite{nanopore}). Other deep-learning architectures can also be investigated, including those employing transformers. Developing new architectures that can be efficiently trained for decoding long codewords is also interesting.

\end{document}